\documentclass{article}
\usepackage[letterpaper,margin=1in]{geometry}

\usepackage{amsmath,amssymb,amsfonts}
\usepackage{algorithmic}
\usepackage{algorithm}
\usepackage{enumitem}
\usepackage{graphicx}
\usepackage[caption=false,font=normalsize,labelfont=sf,textfont=sf]{subfig}
\usepackage{textcomp}
\usepackage{stfloats}
\usepackage{url}
\usepackage{verbatim}
\usepackage{xcolor}
\usepackage{makeidx}

\def\capmystringaux#1#2\relax{\uppercase{#1}\lowercase{#2}}

\newcommand{\newdef}[1]{\textit{#1}\ignorespaces\index{#1}}

\newcommand{\newda}[2]{\textit{#1}\ignorespaces\index{#1}\index{#1|seealso {#2}} (\textit{#2})\index{#2}\index{#2|seealso {#1}}}

\makeindex

\title{Annotated PIM Bibliography\\ Preliminary Version 0}
\author{Peter Kogge}
\date{February 2025}

\begin{document}

\maketitle

\setcounter{tocdepth}{2}
\tableofcontents

\section{Introduction}\label{sec:intro}

\textit{Processing in Memory} (PIM) and similar terms such as \textit{Compute In Memory} (CIM), 
\textit{Logic in Memory} (LIM),
\textit{In Memory Computing} (IMC),
and \textit{Near Memory Computing} (NMC)  have gained attention recently as a  potentially ``revolutionary new'' technique. The truth, however, is that many examples of the technology go back over 60 years. This document attempts to provide an annotated bibliography of PIM technology that attempts to cover the whole time-frame, and is organized to augment a forth-coming article.

\subsection{Organization}

Given the large number of papers, an attempt has been made to group them in an organized and hierarchal fashion. 

The highest grouping is whether the contents of a paper is deemed some sort of ``survey'' of PIM designs or PIM-relevant technologies (Section  \ref{sec:surveys}), a discussion of specific PIM-relevant technologies, but not a ``complete'' system (Section \ref{sec:other}), or a complete system (Sections \ref{sec:studies} through \ref{sec:commercial}). 

Section \ref{sec:unsorted} lists papers that must be sorted and entered into the correct previous sections in a future release of this paper.

In most section, subsections are listed in time-order based on the earliest reference covered in that subsection.

\subsection{Design-specific Papers}

Sections \ref{sec:studies} through \ref{sec:commercial} cover papers that discuss specific PIM designs. Within each section separate subsections address specific designs. All papers that address a specific design are referenced in the same subsection. The subsections are arranged in time order of the earliest paper referenced in the section.

The division into which section a paper goes depends on the degree to which the discussed PIM has reached practice, 
Section \ref{sec:studies} covers designs that had no physical implementations, but were paper or simulation studies only. Section \ref{sec:prototype} covers systems for which some sort of prototype was eventually built, but the implementation was not designed for more than ``lab'' evaluation. Section \ref{sec:commercial} covers papers that
that have reached some level of commercialism, such as multiple copies built, or a system open for multiple users. 

Section \ref{sec:neural} focuses just on PIM-like systems for AI applications, as this area has exploded in recent years.  Papers referenced here may also be referenced in earlier sections, with the emphasis here on what component of AI processing they addressed.

\subsection{Change Log}

Numbers below refer to the document versions.

\begin{enumerate}[noitemsep,nolistsep,leftmargin=*]

\item 1/13/26: First release to establish a link for an upcoming paper. Not all listed references have been documented.

\end{enumerate}

\section{Surveys}\label{sec:surveys}

This section lists surveys of PIM technology, and/or discussion of PIM properties or characteristics.


\begin{itemize}[noitemsep,nolistsep,leftmargin=*]

\item  \cite{2019-Singh-near-memory}

\item \cite{2022-Asufuzzaman-Survey}

\item \cite{2022-Peccerillo-upmem-survey}

\item \cite{2023-Houshmand-BenchmarkingAnalog}

\item \cite{2023-Mutlu-pim-primer}

\item \cite{2024-Khan-survey}

\end{itemize}

\section{Specific System Design Studies}\label{sec:studies}

The following sections discuss PIM designs or studies of  PIM systems targeting specific applications that never reached the prototype or commercial level. General PIM studies can be found in Section \ref{sec:surveys}.

To be added:
\begin{itemize}[noitemsep,nolistsep,leftmargin=*]

\item CAM 1955: \cite{1963-TRW-CAM}

\item SpaceRam 2000: \cite{2000-Margolus-SpaceRAM}

\item Smart Memories 2000: \cite{2000-Mai-SmartMemories}

\item Hybrid SIMD 2022: \cite{2022-Coluccio-Hybrid-SIMD}

\item Transactional: \cite{2007-McDonald-Transactional}

\item PIM for MPI 2003: \cite{2003-Rodrigues-PIM-MPI}

\item 2022 NAND PIM: \cite{2022-Kim-NANDPIM}

\item 2024 Piccolo: \cite{2024-Shin-Piccolo}

\item 2022 RNA Sequence: \cite{2022-Chen-RNASeq}

\item 1997 Why PIM: \cite{1997-Burger-WhyPIM}

\end{itemize}

\subsection{1969: Cellular Logic in Memory}\label{sec:clim}

Very early on, the 2D repetition of identical cells of devices was recognized as having multiple advantages when fabricated with large scale integration. While certainly true when the cells supported memory, it was also true when the cells also included logic. \cite{1969-Kautz-cellular} in particular studied 2D  arrays where each cell contained logic but behaved as a typical addressable memory that could also do sorting, associative lookup, a pushdown memory, and other functions.

Kautz discusses two cell designs, where in both each cell has a connection only to its four neighbors. A register on the equivalent of the row select lines could both send a select signal to all cells in a row, and be set or reset on the basis of signals from the logic in the row of cells.

In  his design, the array acts somewhat like an associative memory in that a data value would be presented as input (notionally ``at the bottom'' of the array, and all rows of cells would compare their value to the input. If they were greater a common row select bit would be set. Then, in parallel all rows whose row select bit is set will accept the value of the row below them take on the value of the row below them. Then the row that is so selected but whose row below is not selected will accept the value presented as input. If the data values are input in this manner one at a time, the result will be that the array contains the data in sorted order, lowest value at the bottom, and highest at the ``top''.

An enhancement includes the ability to include a mask field so that only certain bit positions in each word would participate in the sort. This also permits a \newdef{PLA}-like functionality to be achieved where the array could perform any logical function on the input. The ability to move a selected set of rows up or down also permitted other structures such as pushdown stacks or FIFO queues to be implemented.

\subsection{1970: Compute in Cache}\label{sec:stone}

By 1970 the concept of a \newdef{cache} was just beginning to be considered for inclusion in commercial computers. \cite{1970-Stone-LogicInMemory} made the suggestion to embed compute capabilities in a cache similar to that which was just being employed in the IBM 360 Model 85. In modern terms this cache design was fully associative with 16 ``sector'' entries, each representing 1024 consecutive bytes of memory. An LRU mechanism was used on a miss. The memory system consisted of four interleaved banks of memory, each of which returned 16 bytes on an access, providing a 64B ``block" on each memory access. 16 such blocks were needed to fill a sector on a miss.

Stone's suggestion was to augment the CPU's instruction set with instructions to perform operations on a sector in the cache, and to put logic at the cache to perform them. This logic included a ``tag bit'' with each word in the register. Each such instruction would provide a memory address which would select a sector from memory, and several single-word operands. Suggested operations included:
\begin{itemize}[noitemsep,nolistsep,leftmargin=*]

\item \textit{Search on masked equality}: one operand is a register containing a bit mask, and the other a comparison value. All words in the sector that match have their tag bits set.
\item \textit{Search on masked threshold}: like the above but the test is greater than or equal
\item \textit{Copy Tag Bits}: Move the matching sector's tag bits into specified bits in the operand.
\item \textit{Tag Bit AND}: two operands specif bit positions in each word of the sector that should be ANDed together.

\item \textit{Sector ADD}: Add two sectors together.

\item \textit{Sector Scale}: Multiply each word in a sectore by some value.
\end{itemize}
.

\subsection{*1998: Active Pages}\label{sec:active-pages}

\cite{1998-Oskin-ActivePages}: 1998-Oskin-ActivePages,

\subsection{*1999: Linden DAMM}\label{sec:DAMM}

\cite{1999-Lipovski-DAAM}: 1999-Lipovski-DAAM

\subsection{*2019: DigitalPIM for Data Intensive}\label{sec:digitalpim}

\cite{2019-Imani-DigitalPIM}

\subsection{*2019: DRAM PIM}\label{sec:dram-pim}

\cite{2019-Lee-DRAM-PIM}

\subsection{*2021:MIMHD}\label{sec:mimhd}

\cite{2021-Kazemi-inferencing}

\subsection{*2021: Reconfigurable CIM Array}\label{sec:cim-ml}

\cite{2021-Anni-ML}

\subsection{*2022: MultPIM}\label{sec:multpim}

\cite{2022-Leitersdorf-MultPIM}

\section{Designs That Reached a Prototype Stage}\label{sec:prototype}

The following discuss PIM designs that reached the point where prototype implementations were developed, but not carried to the commercial level. For some of these, there may have been an early study that proposed a design.

\subsection{*1987: J-Machine}\label{sec:jmachine}

\cite{1987-Dally-Jmachine}

\cite{1992-Dally-MDP}: 1992-Dally-MDP

\cite{1993-Noakes-MDP}: 1993-Noakes-MDP

\cite{1990-jmachine-dataflow} 1990-jmachine-dataflow

 \cite{1988-Dally-J-actors} 1988-Dally-J-actors

\subsection{*1992: CAM}\label{sec:storom-cam}

\cite{1992-Stormon-CAM}: 1992-Stormon-CAM

\subsection{1992: RTAIS - Real Time AI Systems}\label{sec:rtais}

The \newda{RTAIS}{Real Time AI System} \cite{1995-Kogge-RTAIS} was a prototype system developed at IBM's Federal Systems Division in Owego, NY in the late 1980s in an era where expert systems were becoming popular, and there was a drive to perform such computations fast enough to be used in an embedded real-time environment. 

As implemented the system was built around seven modules all of which shared a common memory bus. Four modules each held a MIPS R-3000 processor. Two other modules each held 64MB of memory. The final modules termed the \newda{ACP}{Associative Control Processor} was a PIM structure containing memory that was addressable as normal over the shared memory bus, but could also perform ``in-memory'' computations.

The physical memory on the ACP was divided into 30 separate 128KB byte-wide banks of SRAM memory. Each bank had at its row buffer a small 8-bit ALU designed into a FPGA to implement a rich variety of comparison operations.  A small control memory on the module could be loaded with programs to be executed by the memory-embedded processing logic.

Once triggered by a special write to the ACP, a sequence of SIMD accesses and processing would be invoked. This was equivalent to 30 separate but ganged vector operations. 

While numeric operations were possible, the main vector operations were based on comparisons. Each comparison consisted of a reference datum to use in the comparison, a compare operator, and a datum value to to check. If the first pair came from the R3000, the operation was equivalent to a sequential database search, with the database in memory. However, the more important configuration was when the reference datum and the operator were stored in memory, and the compare datum came over the memory bus from the R3000. This case was equivalent to a ``database'' of patterns, where the question to be answered by the overall operation was which pattern did some set of data match. This dual to a database search was at the heart of the programming language \newdef{Linda}, and allowed streams of incoming data to be compared to sets of trigger rules, and alerts generated when there was a match as to which rule was matched.

\subsection{1997: IRAM/VIRAM}\label{sec:IRAM}

The IRAM approach to PIM was to integrate multiple vector processors onto a memory die, and use all the memory bandwidth to perform both SIMD operations across all data returned in a single access, and then follow up with multiple such accesses. The original studies \cite{1997-Patterson-IRAM}, \cite{1997-Kozyrakis-IRAM}, \cite{1999-Patterson-IRAM} postulated a memory die with a conventional core at the center, with multiple vector units connected by crossbars to the memory arrays.

A prototype chip \cite{2004-Gebis-VIRAM} was implemented in 2002 in 180nm CMOS, with 125 million transistors. There were eight banks of 13Mbit DRAM with 256 bit interfaces connected by crossbars to four \newdef{vector lanes}, each implementing a vector core. Each such core contained an 8KB vector register and two arithmetic units, one of which could support single precision floating point. A suite of media-oriented operations were included.
A MIPS M5Ke core served as an on chip control processor. 

At 200MHz power was about 2 watts, and performance against a variety of other embedded processors indicated a very significant gain for VIRAM.

\subsection{*1999: DIVA}\label{sec:diva}

"That project, DIVA, built a 55-million transistor PIM chip that the DIVA team demonstrated in a development system July 31, 2002, at the DARPATech 2002 Symposium in Anaheim California."

\cite{1999-Hall-DIVA}: 1999-Hall-DIVA

\cite{2000-Kang-DIVA}: 2000-Kang-DIVA

\cite{2002-Chiueh-DIVA}: 002-Chiueh-DIVA
\cite{2002-Draper-DIVA}: 2002-Draper-DIVA

\cite{2004-Mediratta-DIVA}: 2004-Mediratta-DIVA

\cite{2004-Mediratta-DIVA-exceptions}: 2004-Mediratta-DIVA-exceptions

\cite{2004-Kwon-DIVA,2006-Barrett-DIVA}: 2004-Kwon-DIVA,2006-Barrett-DIVA

\subsection{*2003: Yukon}\label{sec:yukon}

\cite{2003-Kirsch-Yukon}: 2003-Kirsch-Yukon

\subsection{*2004: PIM Lite}\label{sec:pimlite}

 \cite{2003-Brockman-PIMLite}: 2003-Brockman-PIMLite

 \cite{2003-Brockman-PIMLite-Asm}: 2003-Brockman-PIMLite-Asm
 
 \cite{2003-Brockman-PIMLite-programming}: 2003-Brockman-PIMLite-programming
 
\cite{2004-Brockman-PIMLite}: 2004-Brockman-PIMLite

 \cite{2005-Thoziyoor-PIMLite}: 2005-Thoziyoor-PIMLite
 
 \cite{2007-Li-PimLite}: 2007-Li-PimLite

\subsection{*2023: DIANA}\label{sec:diana}

\cite{2023-Houshmand-DIANA}: 2023-Houshmand-DIANA

\section{Commercial and Near-Commercial Systems}\label{sec:commercial}

The following discuss PIM systems that reached a maturity where a commercial or near-commercial versions were eventually available.

\begin{itemize}[noitemsep,nolistsep,leftmargin=*]

\item Associative Memory: \cite{2006-pagiamtzis-content}






\item MASPAR 1990: \cite{1990-Blank-MASPAR,1990-Maspar-Fortran}

\item Mitsubishi M32: \cite{1996-Shimizu-M32,
1997-Nunomura-M32,
1998-Nunomura-M32}

\item Micron Automata: \cite{2016-Wang-Automata}

\end{itemize}

\subsection{1972: {STARAN}}\label{sec:staran}

In the early 1970s Goodyear Aerospace Corp. delivered a series of computers called STARAN \cite{1974-Batcher-Staran} for use in highly parallel real-time applications such as air traffic control. The unique component of STARAN was a multi-dimensional memory module 
\cite{1977-Batcher-Staran-memory} that could be accessed both by word and bit-slice (i.e. access bit ``i'' of all words simultaneously). 

STARAN's 256-word by 256-bits-per-word memory had fairly conventional but wide for the time (256 bits) read and write busses, along with an address and command bus. However, like an associative memory it also had a Mask bus. This mask could control which bits in an access should be modified on a write. More uniquely, there was a \newdef{flip network} \cite{1976-Batcher-Staran-network} that allowed the bits of the memory to be accessed in many different ways, especially as either conventional ``words'' or as ``bit slices'' with one bit from each word.

Integrated into the memory module was an array of 256 single-bit \newda{Processing Elements}{PEs} that could take the output of the flip network and perform arbitrary boolean operations on them.

A typical STARAN system had four such modules controlled an \newda{associative Processor}{AP}  that had direct access to the registers in the PEs. A conventional computer ran the main program, and accessed the AP as a special purpose peripheral. 

\cite{{1974-Batcher-Staran}} describes multiple bechmarks where the STARAN was literally orders of magnitude faster than conventional processors at the time.

\cite{1974-Davis-staran-sw} 1974-Davis-staran-sw

\subsection{*1972: ICL DAP}\label{sec:DAP}

\cite{1973-Reddaway-DAP}

\subsection{1981: {ASPRO - Airborne Associative Processor}}\label{sec:aspro}

The 1971 \newdef{STARAN} (Section \ref{sec:staran}) was built before VLSI technology was available. Later in the 1970s the same group designed a new version of STARAN using VLSI, called the \newda{Airborne Associative Processor}{ASPRO} \cite{1982-Batcher-ASPRO-MPP}. 

Each memory array module contained sixteen 1K by 4-bit memories and four VLSI logic chips, each containing 32 PEs and a 32-bit flip network. Thus each module held 128 by 4096 bits of memory and 128PEs.
There were 17 such array modules in an ASPRO system (one was a spare).

Again the main application focus was on real-time airborne traffic control.

\subsection{1982: {MPP - Massively Parallel Processor}}\label{sec:mpp}

The 1971 \newdef{STARAN} (Section \ref{sec:staran}) was built before VLSI technology was available. A later version, \newdef{ASPRO} (Section \ref{sec:aspro}) was built in 1981 for the same application space - real-time airborne traffic control. The same group that built these machines was also contracted to design a similar machine for NASA to do image processing. This system \cite{1982-Batcher-ASPRO-MPP} had 16K \newdef{PE}'s arranged in a 128 row X 132 column rectangular array (there wee an extra 4 columns for redundancy.

Each PE was now capable of a richer operation suite, including 32 bit floating point, and had communication links with its four neighbors in a 2D arrangement.

\subsection{1982: GAPP - Geometric Arithmetic Parallel Processor}\label{sec:gapp}

The \newda{Geometric Arithmetic Parallel Processor}{GAPP} \cite{1988-Cloud-GAPP} was a project of the Martin Marietta Electronics Systems Division to develop a massive SIMD system for avionics problems. It used an array of 2 micron CMOS chips, where each chip held an array of cells, each of which in turn contained memory and logic. Systems of up to 83K cells were built.

Each cell contained 128 bits of memory and logic consisting of four 1-bit registers and a 1-bit ALU. A common control path of 13-bits to control the ALU and data movement within the cell, and 7 bits of address for the memory. 

Each chip held 12 by 6 cells. One bit on each cell was arranged into a ``North-South'' shift register, and another allowed ``East-West'' shifts. A third allowed shifts in the ``North'' directions only. A chip ran at 10MHz and dissipated about 1/2 watt.

The processor array was wired as an peripheral to a DEC VAX processor. A single array board contained 88 chips arranged as $48\times 132=6336$ cells.

A separate \newda{GAPP Module Controller}{GMC} sits between the host and the array, and holds the low level routines to perform operations in the array.

A separate study \cite{1988-Signorini-GAPP} investigated extensions that would allow virtual reconfiguration of the calls in a GAPP array, and provide non-local communications between arbitrary cells.

\subsection{1985: CM-1: Connection Machine-1}\label{sec:CM1}

The \newda{Connection Machine - 1}{CM-1} \cite{1988-Tucker-CM1} was a commercialized outgrowth of an early 1980's Ph.D. thesis by Daniel Hillis at MIT which was later published in book form \cite{1989-Hillis-cm}. The first prototype was delivered in 1984, and commercial versions in 1986.

Architecturally it was in the style of the \newdef{STARAN} (Section \ref{sec:staran}), \newdef{DAP} (Section \ref{sec:DAP}), \newdef{ASPRO} (Section \ref{sec:aspro}), and \newdef{MPP} (Section \ref{sec:mpp}). Memory was partitioned into up to 65K partitions of 4K-bits, each of which was associated with a small 1-bit ALU termed a \newda{Processing Element}{PE} that contained a few 1-bit registers. Multiple PE logic was implemented on a sngle VLSI chip.

Each set of 16K PEs was controlled by a\newda{Sequencer} that could broadcast commands simultaneously to all PEs in the group. The commands consisted of multiple addresses into each PEs' memory and an operation to perform on them. The instructions typically took the form of memory-to-memory operations - 2 separately addressed bits of memory would be run through the ALU in the PR, and run back to a possible different bit in memory.

Up to four sequences could be ganged together in a system, each with its own 16K PEs. These Sequences in turn could be controlled by up to four host processors, typically \newdef{DEC VAX}s.

A unique aspect of the CM-1's instruction set (called \newdef{PARIS} \cite{1989-ThinkingMachines-Paris} was that programs could be written for ``virtual CM-1s'' with far more than the physical number of PEs provided in the hardware. At the beginning of a program, the desired number of PEs could be defined, and a block of each physical PE's memory allocated for it. Then when a command to perform operations on each virtual PE, each physical PE would repeat the operation on each virtual copy that it held.

Besides the broadcast bus, the PEs were interconnected by a \newdef{Global OR} (each PE could logically OR a bit into it and all could see the result), a \newdef{NEWS network} (Each PE could transfer data to/from its four neighbors in a 2D array, and twelve connections into a 12-dimensional hypercube network.
This latter network was one of the more unique aspects of the CM-1 design, in that each PE could prepare (in parallel) a message to be delivered to any other PE in the system (with the address of that PE dynamically computable by the PE), launched into the network (in parallel), and delivered (in parallel) to the target PE. This was particularly useful when large numbers of virtual PEs wanted to exchange data.

A variety of programming languages were adapted to run on the machine, including FORTRAN, Lisp, and C.

\cite{1988-Tucker-CM1}

\cite{1989-Kahle-CM1}

\subsection{1988: CM-2: Connection Machine-2}\label{sec:CM2}

The \newda{Connection Machine-2}{CM-2} was an upgrade to the \newdef{CM-1} (Section \ref{sec:CM1}) that was faster, with more capacity, and better reliability. It also added an I/O system and a hardware floating-point capability. 

Each \newdef{PE} had 16 times that of the CM-1 (64K-bits), with the logic for 16 PEs packaged on a single VLSI chip. Every two PE chips (with 32 PEs) then had connections to a special floating point memory interface and a separate floating point accelerator. Thus 32 PEs ``looked like memory'' to the floating point chip.

Each matrix board then held 32 PEs, and 16 each of the floating point memory interface and accelerator chips, for a total of 512 PEs and 32M-bits,

The physical design of the CM-2 ``package'' was a deliberate attempt to introduce some aesthetics that would stress its novel architecture \cite{1994-Theil-CM2}. Its iconic front panel was then made famous in the movie ``Jurassic Park.''

The performance of the CM-2 \cite{1988-Sato-CM2-benchmarking} over a wide range of applications was quite remarkable for its time.

\cite{1990-TMC-CM2-sw} 1990-TMC-CM2-sw

\subsection{1992: EXECUBE}\label{sec:execube}

\newdef{EXECUBE} was the name of a project from IBM's Federal Systems Division in Owego, NY to develop a single chip types that held both memory and compute, and could be scaled to large parallel compute systems. The goal was to find ways to effectively utilize the bandwidth provided by a DRAM array without discarding any of it to go off chip. The successful chip  design took a then common 4Mbit DRAM part, partitioned the sub-arrays into 8 sections, and integrated compute logic into the sense amps and row buffers of each partition. 

Each 64KB partition had its own 16-bit core next to it. The 8 cores in each chip could be run in MIMD (all cores run independent programs) or SIMD (all cores execute instructions broadcast from outside the chip, or a dynamic mix of the two.

Each core had two kinds of interfaces. First was a common \newdef{broadcast bus} that all cores monitored, and that allowed an external processor to transmit commands the arbitrary combinations of the cores. These commands could be MIMD program setup commends targeting specific partitions or SIMD instructions to be executed by a subset of cores synchronously and concurrently.

Second were four full-duplex inter-partition links associated with each partition. Three of each partition's links were to other on-chip partitions, forming a complete hypercube. The fourth went off chip, and could be connected directly to any port on any other EXECUBE chip. Each of these links supported a DMA-like protocol in both directions. Communication could be in ``packet mode'' where data from one partition's memory could be moved to the memory of the partition on the other side, or in a ``circuit-switched'' mode, where partitions could be commanded to route data streams coming in one link to exit another, without utilizing the on-partition memory.

Working chips were first produced in 1992, and systems of up to 64 of them integrated into an IBM RISC-6000 workstation. All chips had their broadcast bus ports tied together to a \newdef{Cluster Controller} implemented as an I/O card on the host the host, and the external links joined to form a 4D hypercube.

\cite{1994-kogge-ICPP} gave the first overview of the chip and systems built from the chip. \cite{1995-kogge-Execube-system} compares the chip to other systems at the time. \cite{1996-Sunaga-Execube} discusses in detail the technology used to implement the chip. \cite{1997-Sheliga-Execube} discusses algorithms for image compression that might use such chips. The general architecture drove one of the three 1995 proposed architectures for petascale computers \cite{1995-Sterling-petaflops}.

\subsection{*1993: Cray T3D}\label{sec:t3d}

\cite{1993-Kessler-T3D}

\subsection{1995: Terasys}\label{terasys}

The goal of the \newdef{Terasys} project \cite{1995-Gokhale-Terasys} was to design a standard memory chip that included a 1-bit ALU under each memory bit column that could perform SIMD operations on the data returned from a memory access. The design was reduced to practice in a 128Kbit SRAM with 2K 64-bit words, and 64 1-bit ALUs. The interface was a standard SRAM protocol with a 4-bit data bus.

While the target system was to be memory for the Cray-3 supercomputer, a half dozen SUN Sparcstation-2 workstations were constructed, each with 8MB of the above chip.

The ALUs included three 1-bit registers that could hold reads from memory, or transfers from neighboring ALUs. A Global OR combines a signal from each  A Partitioned OR provided a similar signal on a power of 2 subset of a system's ALUs. A Parallel Prefix Network allowed signals to be sent from each ALU to one whose index was a power of two different (this supported log time additions).

The interface to the host processor was a standard address-command pair. If the highest order address bit was a 0, the address selected a word from the memory and the command specified a conventional read or write. If the highest order address bit was a 1, the command was re-interpreted as an ALU command where the lower order address bits were re-interpreted into an index into an on-chip table of micro-coded instructions.

A variant of C was developed as a programming language that included a variety of parallel extensions.

\cite{1995-Gokhale-Terasys} describes the chip in detail.

\subsection{1997: Mitsubishi M32R/D}\label{sec:M32}

The embedding of processor cores into chips designed with a DRAM-centered technology that started with EXECUBE (Section \ref{sec:execube}) continued with the Mitsubishi \newdef{M32R/D} \cite{1997-Nunomura-M32}. This chip supported a 2 MB DRAM array where the row buffers formed a cache for a more 66MHz robust 32-bit RISC core implemented by a five-stage execution pipeline. The targeted applications were portable multimedia systems where embedded power was important. An  external bus allowed access to external memory, devices, and peripherals.

\cite{1998-Nunomura-M32} provides performance details on a variety of multimedia kernels.

A second generation design \cite{1998-Shimizu-M32RxD} \newdef{M32Rx/D} used 250nm CMOS with only 3 layers of metal on a 100 $mm^2$ die. It doubled the memory capacity, supported a dual issue core at 100MHz, and added \newda{Digital Signal Processing}{DSP} instructions. Further the chip can look like a memory chip to outside devices, with the on-chip core running asynchronously. Thus the chip could look like a single-chip processor with memory, intelligent memory, and/or a node in a multi-processor.

\subsection{2004: Migrating Threads}\label{sec:emu}

Modern parallel computers are almost uniformly identified as \newdef{distributed memory systems} which consist of sets of \newdef{node}s, each of which holds both conventional processors and memory. A core in a node may only have conventional load/store access to memory in the same node. If a program running in a core wishes to access memory in some other node, it must first (in software) identify what node holds the addressed object, then (in software) create a message to other processors in that node, (in software) give that message to a \newda{network interface controller}{NIC} on its node. The NIC then injects the message into a network that routes the message to a NIC on the targeted node, where software on the target node must again be invoked to read, interpret, and then perform the desired operation (all in software) against the proper memory. Software on the target node must then prepare a response message to be injected into the network to go back the other way.

Starting with \newdef{PIM Lite}, Section \ref{sec:pimlite}, an alternative concept was pursued where the name of a physical ``node,'' or even which core within a node, was of no relevance to a software program running in some core. Instead all memory in all nodes resided in a common \newda{Partitioned Global Address Space}{PGAS}. If a thread executing in some core then makes an access to a memory address that is NOT accessible from the current core, \underline{hardware} not software, detects this. Hardware (not software) then suspends the thread, packages the thread state, and injects it into the network for delivery to the correct physical node. At the other end, again hardware (not software) unpacks the thread state and inserts it into any core in the node that has access to the desired memory address. The thread then continues execution, with no need to know that it had moved from one physical node to another. The ``names'' of physical nodes and cores become irrelevant. All that matters is the memory.

\cite{2004-kogge-migration} formally identified this style of architecture. A startup, \newdef{Emu LLC.} then implemented a proof of concept called \newdef{SPIDERS} using logic implemented in FPGAs. A conventional PC served as a host processor that could inject initial threads into such a system to perform computation ``at a particular memory location.'' Such threads could spawn other child threads that could migrate independently. Programs could be stored either in memory (replicated across nodes) or in some cases embedded in a thread's state. For the most part, program implemented functions to be executed in parallel against objects stored in memory.

A later prototype, called a \newdef{Chick} \cite{2016-Dysart-chick}, supported 8 physical nodes, each with 8 multi-threaded cores and 8 physical banks of memory. Separate conventional processors handled connections to the outside world, and again could inject parent threads into the system to initiate computation. In addition, a separate, ultralight, thread was defined that could be spawned by single instructions in a regular thread, migrate to a target memory, and be executed in the memory controller. Such near-memory computations were invaluable for implementing remote atomic operations and the like. Unlike SPIDERS, the architecture of the Chick was mature enough to run complete programs, and not just functions.

An even more robust implementation, called \newdef{Pathfinder}, \cite{2022-Page-Pathfinder} scaled beyond 8 nodes, and had enhanced functionality, especially in ultralight threads that are executed in the memory controllers. One implementation housed in Georgia Tech's CRNCH center \cite{crnch-pathfinder} has 32 nodes, each with 16 banks of memory and 16 multi-threaded cores.

The Chick and Pathfinder systems were both programmed in a version of the language \newdef{Cilk}.

\cite{2025-kogge-migrating} provides a lengthy history of this style of architecture, with references to multiple algorithms which exhibit enhanced scalability over conventional parallel systems.

\subsection{*2011: Hybrid Memory Cube}\label{sec:hmc}

A prime example of how system architecture can level such technology was Micron's Hybrid Memory Cube (HMC) \cite{2011-Pawlowski-HMC}. This was a product where the bottom die was all logic, and supported both memory controllers, multiple network ports, and message routers and local processors. Stacks of DRAM were on top of this die, with multiple ports down to the logic. Large numbers of such stacks could be interconnected into a sea of memory, and multiple host processors could inject messages into the array to access arbitrary memory. A set of operations akin to what was implemented in the T3D are supported in the logic chip's memory controllers.

\subsection{*2014: {Micron Automata}}\label{sec:automata}

\cite{2016-Wang-Automata}

\cite{2014-Dlugosch-automata}

\subsection{*2024: UPMEM}\label{sec:upmem}

\cite{2024-ortega-PIM-AI}: 2024-ortega-PIM-AI

\cite{2022-Das-UPMEM}: 2022-Das-UPMEM

SDK: \url{https://sdk.upmem.com/2021.3.0/}

\cite{2023-upmem-abumpimp}: 2023-upmem-abumpim

\section{PIM Studies for Neuromorphic Applications}\label{sec:neural}

The following are studies only, as in Section \ref{sec:studies}, but with a specific target of neuromorphic (AI) applications. Because of the large numbers of such studies using different technologies, the studies are divided by technology usage.

\subsection{Neuromorphic PIMs using conventional technology}\label{sec:conventional}

\cite{2019-Gupta-NNPIM}: 2019-Gupta-NNPIM

\cite{2019-Kwon-VehicularAI}: 2019-Kwon-VehicularAI

\cite{2019-Zhao-BNN-PIMs}: 2019-Zhao-BNN-PIMs

\cite{2019-Imani-FloatPIM}: 2019-Imani-FloatPIM

\cite{2021-Roy-PIM-DRAM}: 2021-Roy-PIM-DRAM

\cite{2021-Zhou-PIM-DL}: 2021-Zhou-PIM-DL compiler for PIMs

\cite{2022-Ansari-SPP2D}: 2022-Ansari-SPP2D

\cite{2022-Sutradhar-LUT-PIM}: 2022-Sutradhar-LUT-PIM

\cite{2022-Mandal-COIN}: 2022-Mandal-COIN 

\cite{2022-Herzog-TPU4}: 2022-Herzog-TPU4

\cite{2023-Kiningham-GRIP}: 2023-Kiningham-GRIP

\cite{2023-Singh-PARAG}: 2023-Singh-PARAG

\cite{2023-Deepa-digital-CIM}: 2023-Deepa-digital-CIM

\cite{2023-Wu-DE-C3}: 2023-Wu-DE-C3

\cite{2023-Zhao-2T-DRAM}: 2023-Zhao-2T-DRAM

\cite{2023-Jeong-Ternary}: 2023-Jeong-Ternary

\cite{2023-Fang-NN-Rel-PSP}: 2023-Fang-NN-Rel-PSP

\cite{2024-Afifi-Artemis}: 2024-Afifi-Artemis

\cite{2025-Wang-Fast-OverlaPIM}: 2025-Wang-Fast-OverlaPIM

\cite{2025-Wang-TensorCIM}: 2025-Wang-TensorCIM

\cite{2023-Smagulova-ReNeural}: 2023-Smagulova-ReNeural

\subsection{Neuromorphic PIMs using Resistive RAM}\label{sec:ReRAM}

\cite{2011-Yu-ReRAM}: 2011-Yu-ReRAM

\cite{2016-Chi-Prime}:2016-Chi-Prime

\cite{2018-Long-RRAM-RNN}: 2018-Long-RRAM-RNN 

\cite{2019-Agrawal-SPARE}: 2019-Agrawal-SPARE

\cite{2019-Peng-RRAM-RNN}: 2019-Peng-RRAM-RNN

\cite{2019-Long-FeFET-DNN}: 2019-Long-FeFET-DNN

\cite{2019-Long-ReRAM-DNN}: 2019-Long-ReRAM-DNN

\cite{2020-Yang-ReTransformer}: 2020-Yang-ReTransformer

\cite{2020-WAN-NN-RRAM}: 2020-WAN-NN-RRAM

\cite{2021-Fei-XB-SIM}: 2021-Fei-XB-SIM

\cite{2022-Kubendran-ReRAM}: 2022-Kubendran-ReRAM

\cite{2022-Zhang-Re-FeMAT}: 2022-Zhang-Re-FeMAT

\cite{2023-Zhang-FeFET-CIM}: 2023-Zhang-FeFET-CIM

\cite{2023-Kaushik-MRAM-GAN}: 2023-Kaushik-MRAM-GAN

\cite{2023-Song-Refloat}: 2023-Song-Refloat

\cite{2024-Chen-PointCIM} : 2024-Chen-PointCIM

\cite{2024-Chen-NN-RRAM}: 2024-Chen-NN-RRAM

\subsection{Neuromorphic PIMs using other or mixed technology}\label{sec:other-tech}

\cite{2020-Choi-Flash-BNN}: 2020-Choi-Flash-BNN

\cite{2021-Samiee-S-LIMXNN}: 2021-Samiee-S-LIMXNN

\cite{2022-Wu-NN}: 2022-Wu-NN

\cite{2023-Nasab-MTJ-CNTFET}: 2023-Nasab-MTJ-CNTFET

\cite{2024-Sunny-Optical-PIM}: 2024-Sunny-Optical-PIM

\section{Other}\label{sec:other}

\subsection{*2016: Compute Express Link}\label{sec:capi}

\cite{2016-Hofstee-CAPI}

\cite{2025-CXL-cxl}

\cite{2023-Huang-ReRAM}

\subsection{2024: FastOverlaPIM}

Fast-OverlaPIM \cite{2024-Wang-FoverlaPIM} is a framework designated to identify sections of networks which can utilize overlapping execution of consecutive layers specifically with PIM applications in order to improve runtime performance. This work acts as the follow on to OverlaPIM \cite{2023-Zhou-overlapim}, which is one of, if not, the first works to include the concept of computational overlap during the mapping process. OverlaPIM relies upon decomposing inputs, outputs, and network weights into smaller data spaces consistent with demands of each memory element at a given time step, leading to an expensive bottleneck of evaluating all the data spaces between consecutive layers in order to result in the best performance. To remediate, FastoverlaPIM features a new interface for configuring both hardware and optimization procedures, along with a PIM performance model that accurately evaluates mapping efficiency. To enhance data organization and optimization, it includes a fine-grained data space generation algorithm and a computational analysis algorithm that resolves data dependencies across layers to improve overlapping performance. Additionally, a transformation algorithm expands search capabilities for DNN optimization.

\section{To Be Sorted}\label{sec:unsorted}

The following papers have been identified as PIM-related but they have not yet been reviewed enough to place summaries into the proper subsection.

\begin{itemize}[noitemsep,nolistsep,leftmargin=*]


\item Computational RAM \cite{1992-Elliot-CompRAM}

\item \cite{1996-Nowatzyk-wall}: 1996-Nowatzyk-wall

\item \cite{1997-Burger-WhyPIM}: 1997-Burger-WhyPIM

\item 1999 Media: \cite{1998-Rixner-MediaPIM}

\item VSIA \cite{1999-Birnbaum-VSIA}

\item 2017 RSFQ \cite{2017-Sato-RSFQ}

\item \cite{2018-Mohammedali-acc-in-cloud}: 
2018-Mohammedali-acc-in-cloud

\item 2019 Spintronics CRAM: \cite{2019-Zabihi-Spintronics}

\item \cite{2019-Angizi-bitwise}: 2019-Angizi-bitwise

\item \cite{2019-Peng-Weight-Mapping}: 2019-Peng-Weight-Mapping weight mapping

\item \cite{2019-Angizi-DNN-survey}: 2019-Angizi-DNN-survey self survey

\item \cite{2020-Krishnan-area-opt}: 2020-Krishnan-area-opt area optimization

\item \cite{2022-Wu-new-tech}: 2022-Wu-new-tech comparison

\item \cite{2023-Ding-BNN}: 2023-Ding-BNN binary NN on different arrays

\item \cite{2023-Sayed-BNN-survey}: 2023-Sayed-BNN-survey

\item \cite{2024-PIMSIM-NN}: 2024-PIMSIM-NN

\item \cite{2024-Andrulis-CIMLoop}: 2024-Andrulis-CIMLoop

\item \cite{1992-Eicken-Active-Messages} 1992-Eicken-Active-Messages

\item \cite{1973-Hewitt-actor} 1973-Hewitt-actor

\item \cite{2008-Nickolls-cuda} 2008-Nickolls-cuda

\item \cite{2019-Kogge-pvn-semantics} 2019-Kogge-pvn-semantics

\item \cite{2023-Paul-actor-pgas} 2023-Paul-actor-pgas
\end{itemize}

\bibliographystyle{apalike}
\bibliography{pim,bib}

@INPROCEEDINGS{2022-Page-Pathfinder,
  author={Page, Brian A. and Kogge, Peter},
  booktitle={2022 IEEE/ACM Workshop on Irregular Applications: Architectures and Algorithms (IA3)}, 
  title={The Evolution of a New Model of Computation}, 
  year={2022},
  volume={},
  number={},
  pages={9-18},
  doi={10.1109/IA356718.2022.00008}}

@misc{crnch-pathfinder, 
title={{Near Memory and In-Memory}}, 
howpublished = {\url{https://crnch-rg.cc.gatech.edu/near-memory-and-in-memory/}}, 
journal={College of Computing}, 
publisher={Georgia Tech}, 
author={Young, J.}, 
year={2022},}

@STRING{isscc = {Int. Solid State Circuits Conf.}}

@INPROCEEDINGS{2016-Dysart-chick,  
author={Dysart, Timothy and Kogge, Peter and Deneroff, Martin and Bovell, Eric and Briggs, Preston and Brockman, Jay and Jacobsen, Kenneth and Juan, Yujen and Kuntz, Shannon and Lethin, Richard and McMahon, Janice and Pawar, Chandra and Perrigo, Martin and Rucker, Sarah and Ruttenberg, John and Ruttenberg, Max and Stein, Steve},  booktitle={2016 6th Workshop on Irregular Applications: Architecture and Algorithms (IA3)}, title={Highly Scalable Near Memory Processing with Migrating Threads on the Emu System Architecture}, year={2016},  volume={},  number={},  pages={2-9},  doi={10.1109/IA3.2016.007}
}

@INPROCEEDINGS{2004-kogge-migration,  
author={Kogge, P.M.},  
booktitle={Innovative Architecture for Future Generation High-Performance Processors and Systems (IWIA'04)},   
title={Of Piglets and Threadlets: Architectures for Self-Contained, Mobile, Memory Programming},  
year={2004},  volume={},  number={},  pages={130-138},  
doi={10.1109/IWIA.2004.10005}}

@article{2022-Herzog-TPU4,
author = {Herzog, Benedict and Reif, Stefan and Hemp, Judith and H\"{o}nig, Timo and Schr\"{o}der-Preikschat, Wolfgang},
title = {Resource-demand Estimation for Edge Tensor Processing Units},
year = {2022},
issue_date = {September 2022},
publisher = {Association for Computing Machinery},
address = {New York, NY, USA},
volume = {21},
number = {5},
issn = {1539-9087},
url = {https://doi.org/10.1145/3520132},
doi = {10.1145/3520132},
journal = {ACM Trans. Embed. Comput. Syst.},
month = oct,
articleno = {58},
numpages = {24}
}

@INPROCEEDINGS{2023-Huang-ReRAM,
  author={Huang, Yao-Hung and Hsieh, Yu-Cheng and Lin, Yu-Cheng and Chih, Yue-Der and Wang, Eric and Chang, Jonathan and King, Ya-Chin and Lin, Chrong Jung},
  booktitle={2023 IEEE Symposium on VLSI Technology and Circuits (VLSI Technology and Circuits)}, 
  title={High Density Embedded 3D Stackable Via RRAM in Advanced MCU Applications}, 
  year={2023},
  volume={},
  number={},
  pages={1-2},
  doi={10.23919/VLSITechnologyandCir57934.2023.10185230},
  ISSN={2158-9682},
  month={June},}

@article{2019-Singh-near-memory,
title = {Near-memory computing: Past, present, and future},
journal = {Microprocessors and Microsystems},
volume = {71},
pages = {102868},
year = {2019},
issn = {0141-9331},
doi = {https://doi.org/10.1016/j.micpro.2019.102868},
url = {https://www.sciencedirect.com/science/article/pii/S0141933119300389},
author = {Gagandeep Singh and Lorenzo Chelini and Stefano Corda and Ahsan Javed Awan and Sander Stuijk and Roel Jordans and Henk Corporaal and Albert-Jan Boonstra}
}

@article{2006-pagiamtzis-content,
  title={Content-addressable memory (CAM) circuits and architectures: A tutorial and survey},
  author={Pagiamtzis, Kostas and Sheikholeslami, Ali},
  journal={IEEE journal of solid-state circuits},
  volume={41},
  number={3},
  pages={712--727},
  year={2006},
  publisher={IEEE}
}

@ARTICLE{1992-Stormon-CAM,
  author={Stormon, C.D. and Troullinos, N.B. and Saleh, E.M. and Chavan, A.V. and Brule, M.R. and Oldfield, J.V.},
  journal={IEEE Micro}, 
  title={A general-purpose CMOS associative processor IC and system}, 
  year={1992},
  volume={12},
  number={6},
  pages={68-78},
  doi={10.1109/40.180249}}

@TECHREPORT{1963-TRW-CAM,
  author = {Dudley Allen Buck },
  title = {First interim report on optimum utilization of computers and computing techniques in shipboard weapons control systems},
  year = {1963},
 url = {\url{https://citeseerx.ist.psu.edu/document?repid=rep1&type=pdf&doi=c4bbfa7e7519337824709da275cf3f6c26314475}},
  institution = {TRW},
 number = {BuWeps-Project RM1004 M88-3U1}
}

@INPROCEEDINGS{2016-Wang-Automata,
  author={Wang, Ke and Angstadt, Kevin and Bo, Chunkun and Brunelle, Nathan and Sadredini, Elaheh and Tracy, Tommy and Wadden, Jack and Stan, Mircea and Skadron, Kevin},
  booktitle={2016 International Conference on Hardware/Software Codesign and System Synthesis (CODES+ISSS)}, 
  title={An overview of Micron's Automata processor}, 
  year={2016},
  volume={},
  number={},
  pages={1-3},
  doi={}}

@ARTICLE{2014-Dlugosch-automata,
  author={Dlugosch, Paul and Brown, Dave and Glendenning, Paul and Leventhal, Michael and Noyes, Harold},
  journal={IEEE Transactions on Parallel and Distributed Systems}, 
  title={An Efficient and Scalable Semiconductor Architecture for Parallel Automata Processing}, 
  year={2014},
  volume={25},
  number={12},
  pages={3088-3098},
  doi={10.1109/TPDS.2014.8}}

@ARTICLE{2016-Hofstee-CAPI,
  author={Peter Hofstee},
  journal={23rd Reconfigurable Architectures Workshop  at 30th IEEE International Parallel and
Distributed Processing Symposium}, 
  title={Keynote: Reconfigurable Accelerators for Big Data and Cloud}, 
  year={2016},
  volume={},
  number={},
  pages={},
howpublished = {\url{https://raw.necst.it/2016/RAW-keynote-Hofstee-final.pdf}}
}

@misc{2025-CXL-cxl,
  author       = {CXL},
  title        = {{CXL 3.2 Specification}},
  howpublished = {\url{https://computeexpresslink.org/}},
  month        = "",
  year         = {2025},
  note         = ""
}

@ARTICLE{1989-Kahle-CM1,
  author={Kahle, B.A. and Hillis, W.D.},
  journal={IEEE Transactions on Systems, Man, and Cybernetics}, 
  title={The Connection Machine model CM-1 architecture}, 
  year={1989},
  volume={19},
  number={4},
  pages={707-713},
  doi={10.1109/21.35335}}

@ARTICLE{1988-Tucker-CM1,
  author={Tucker, L.W. and Robertson, G.G.},
  journal={Computer}, 
  title={Architecture and applications of the Connection Machine}, 
  year={1988},
  volume={21},
  number={8},
  pages={26-38},
  doi={10.1109/2.74}}

@techreport{1989-ThinkingMachines-Paris,
  author      = "Thinking Machines",
  title       = "Paris Reference Manual",
  institution = "Thinking Machines",
  year        = "1989",
  type        = "",
  number      = "Tech Report Version 6.0",
  address     = "",
  month       = "",
  howpubpished = {\url{https://bitsavers.computerhistory.org/pdf/thinkingMachines/CM2/Paris_Reference_Manual_Verison_6.0_Feb1991.pdf}},
}

@INPROCEEDINGS{1988-Sato-CM2-benchmarking,
  author={Sato, R.K. and Swarztrauber, P.N.},
  booktitle={Supercomputing '88:Proceedings of the 1988 ACM/IEEE Conference on Supercomputing, Vol. I}, 
  title={Benchmarking the Connection Machine 2}, 
  year={1988},
  volume={},
  number={},
  pages={304-309},
  doi={10.1109/SUPERC.1988.44667}}

@book{1989-Hillis-cm,
author = {Hillis, W. Daniel},
title = {The Connection Machine},
year = {1989},
isbn = {0262081571},
publisher = {MIT Press},
address = {Cambridge, MA, USA}
}

@INPROCEEDINGS{1994-Theil-CM2,
  author={Thiel, Tamiko},
  booktitle={Design Issues}, 
  title={The Design of the Connection Machin}, 
  year={1994},
  volume={10},
  number={1},
  pages={5–18},
  doi={10.2307/1511650}}

@INPROCEEDINGS{1999-Lipovski-DAAM,
  author={Lipovski, G.J. and Yu, C.},
  booktitle={Records of the 1999 IEEE International Workshop on Memory Technology, Design and Testing}, 
  title={The dynamic associative access memory chip and its application to SIMD processing and full-text database retrieval}, 
  year={1999},
  volume={},
  number={},
  pages={24-31},
  doi={10.1109/MTDT.1999.782680}}

@inproceedings{1973-Reddaway-DAP,
author = {Reddaway, S. F.},
title = {DAP—a distributed array processor},
year = {1973},
isbn = {9781450374286},
publisher = {Association for Computing Machinery},
address = {New York, NY, USA},
url = {https://doi.org/10.1145/800123.803971},
doi = {10.1145/800123.803971},
booktitle = {Proceedings of the 1st Annual Symposium on Computer Architecture},
pages = {61–65},
numpages = {5},
series = {ISCA '73}
}

@INPROCEEDINGS{1999-Hall-DIVA,
  author={Hall, M. and Kogge, P. and Koller, J. and Diniz, P. and Chame, J. and Draper, J. and LaCoss, J. and Granacki, J. and Brockman, J. and Srivastava, A. and Athas, W. and Freeh, V. and Jaewook Shin and Joonseok Park},
  booktitle={SC '99: Proceedings of the 1999 ACM/IEEE Conference on Supercomputing}, 
  title={Mapping Irregular Applications to DIVA, a PIM-based Data-Intensive Architecture}, 
  year={1999},
  volume={},
  number={},
  pages={57-57},
  doi={}}

@inproceedings{2002-Draper-DIVA,
author = {Draper, Jeff and Chame, Jacqueline and Hall, Mary and Steele, Craig and Barrett, Tim and LaCoss, Jeff and Granacki, John and Shin, Jaewook and Chen, Chun and Kang, Chang Woo and Kim, Ihn and Daglikoca, Gokhan},
title = {The architecture of the DIVA processing-in-memory chip},
year = {2002},
isbn = {1581134835},
publisher = {Association for Computing Machinery},
address = {New York, NY, USA},
url = {https://doi.org/10.1145/514191.514197},
doi = {10.1145/514191.514197},
booktitle = {Proceedings of the 16th International Conference on Supercomputing},
pages = {14–25},
numpages = {12},
location = {New York, New York, USA},
series = {ICS '02}
}

@INPROCEEDINGS{2004-Mediratta-DIVA,
  author={Mediratta, S. and Sondeen, J. and Draper, J.},
  booktitle={17th International Conference on VLSI Design. Proceedings.}, 
  title={An area-efficient router for the Data-Intensive Architecture (DIVA) system}, 
  year={2004},
  volume={},
  number={},
  pages={863-868},
  doi={10.1109/ICVD.2004.1261039}}

@INPROCEEDINGS{2002-Chiueh-DIVA,
  author={Herming Chiueh and Draper, J. and Mediratta, S. and Sondeen, J.},
  booktitle={Proceedings of the 28th European Solid-State Circuits Conference}, 
  title={The address translation unit of the data-intensive architecture (DIVA) system}, 
  year={2002},
  volume={},
  number={},
  pages={767-770},
  doi={}}

@INPROCEEDINGS{2000-Kang-DIVA,
  author={Chang Woo Kang and Draper, J.},
  booktitle={Proceedings of the 43rd IEEE Midwest Symposium on Circuits and Systems (Cat.No.CH37144)}, 
  title={A fast, simple router for the Data-Intensive Architecture (DIVA) system}, 
  year={2000},
  volume={1},
  number={},
  pages={188-192 vol.1},
  doi={10.1109/MWSCAS.2000.951617}}

@INPROCEEDINGS{2004-Mediratta-DIVA-exceptions,
  author={Mediratta, S. and Steele, C. and Singh, R. and Sondeen, J. and Draper, J.},
  booktitle={The 2004 47th Midwest Symposium on Circuits and Systems, 2004. MWSCAS 2004.}, 
  title={A 0.18/spl mu/m CMOS implementation of an area efficient precise exception handling unit for processing-in-memory systems}, 
  year={2004},
  volume={3},
  number={},
  pages={455-458},
  doi={10.1109/MWSCAS.2004.1354393}}

@INPROCEEDINGS{2006-Barrett-DIVA,
  author={Barrett, T. and Sumit Mediratta and Taek-Jun Kwon and Ravinder Singh and Sachit Chandra and Sondeen, J. and Draper, J.},
  booktitle={2006 IEEE International Symposium on Circuits and Systems (ISCAS)}, 
  title={A double-data rate (DDR) processing-in-memory (PIM) device with wideword floating-point capability}, 
  year={2006},
  volume={},
  number={},
  pages={4 pp.-},
  doi={10.1109/ISCAS.2006.1692989}}

@INPROCEEDINGS{2004-Kwon-DIVA,
  author={Taek-Jun Kwon and Joong-Seok Moon and Sondeen, J. and Draper, J.},
  booktitle={2004 IEEE International Symposium on Circuits and Systems (ISCAS)}, 
  title={A 0.18 /spl mu/m implementation of a floating-point unit for a processing-in-memory system}, 
  year={2004},
  volume={2},
  number={},
  pages={II-453},
  doi={10.1109/ISCAS.2004.1329306}}

@ARTICLE{1994-kogge-ICPP,
  author = {Peter M. Kogge},
  title = {{EXECUBE}-A New Architecture for Scaleable MPPs},
  journal = {Int. Conf. on Parallel Processing},
  year = {1994},
  volume = {1},
  pages = {77-84},
  month = {Aug.},
  doi = {10.1109/ICPP.1994.108}
}

@ARTICLE{1995-kogge-Execube-system,
  author = {Peter M. Kogge and Sunaga, T. and Miyataka, H. and Kitamura, K. and
	Retter, E.},
  title = {Combined {DRAM} and logic chip for massively parallel systems},
  journal = {Proc. Sixteenth Conf. on Advanced Research in VLSI},
  year = {27-29 Mar 1995}
}

@ARTICLE{1996-Sunaga-Execube,
  author = {T. Sunaga and H. Miyatake and K. Kitamura and Peter M. Kogge and
	E. Retter},
  title = {A parallel processing chip with embedded {DRAM} macros},
  journal = {IEEE J. of Solid-State Circuits},
  year = {1996},
  volume = {31},
  pages = {1556-1559},
  number = {10},
  month = {Oct},
  doi = {10.1109/4.540068},
  issn = {0018-9200}
}

@inproceedings{1997-Sheliga-Execube,
  author={Sheliga, M. and Kogge, P.M. and Hsing-Mean Sha, E.},
  booktitle={Proceedings of the IEEE 1997 National Aerospace and Electronics Conference. NAECON 1997}, 
  title={Hardware/software codesign for video compression using the EXECUBE processing array}, 
  year={1997},
  volume={1},
  number={},
  pages={279-286 vol.1},
  doi={10.1109/NAECON.1997.618091}}

@book{1995-Sterling-petaflops,
author = {Sterling, Thomas and Messina, Paul and Smith, Paul H.},
title = {Enabling technologies for petaflops computing},
year = {1995},
isbn = {0262691760},
publisher = {MIT Press},
address = {Cambridge, MA, USA}
}

@INPROCEEDINGS{1988-Cloud-GAPP,
  author={Cloud, E.L.},
  booktitle={Proceedings., 2nd Symposium on the Frontiers of Massively Parallel Computation}, 
  title={The geometric arithmetic parallel processor}, 
  year={1988},
  volume={},
  number={},
  pages={373-381},
  doi={10.1109/FMPC.1988.47455}}

@INPROCEEDINGS{1988-Signorini-GAPP,
  author={Signorini, J.},
  booktitle={Proceedings of the International Workshop on Artificial Intelligence for Industrial Applications}, 
  title={A routing model for the NCR GAPP}, 
  year={1988},
  volume={},
  number={},
  pages={517-521},
  doi={10.1109/AIIA.1988.13341}}

@ARTICLE{2011-Pawlowski-HMC,
  author={J. Thomas Pawlowski},
  journal={Hot Chips 23}, 
  title={Hybrid Memory Cube (HMC)}, 
  year={2011},
  volume={},
  number={},
  pages={},
}

@ARTICLE{1997-Kozyrakis-IRAM,
  author={Kozyrakis, C.E. and Perissakis, S. and Patterson, D. and Anderson, T. and Asanovic, K. and Cardwell, N. and Fromm, R. and Golbus, J. and Gribstad, B. and Keeton, K. and Thomas, R. and Treuhaft, N. and Yelick, K.},
  journal={Computer}, 
  title={Scalable processors in the billion-transistor era: IRAM}, 
  year={1997},
  volume={30},
  number={9},
  pages={75-78},
  doi={10.1109/2.612252}}

@INPROCEEDINGS{1997-Patterson-IRAM,
  author={Patterson, D. and Anderson, T. and Cardwell, N. and Fromm, R. and Keeton, K. and Kozyrakis, C. and Thomas, R. and Yelick, K.},
  booktitle={1997 IEEE International Solids-State Circuits Conference. Digest of Technical Papers}, 
  title={Intelligent RAM (IRAM): chips that remember and compute}, 
  year={1997},
  volume={},
  number={},
  pages={224-225},
  doi={10.1109/ISSCC.1997.585348}}

@INPROCEEDINGS{2004-Gebis-VIRAM,
  author={Joseph Gebis and Sam Williams and David
Patterson},
  booktitle={2004 Design Automation Conference (DAC)}, 
  title={VIRAM1: A Media­Oriented Vector Processor with Embedded DRAM}, 
  year={2004},
  volume={},
  number={},
  pages={},
  doi={}}

@INPROCEEDINGS{1999-Patterson-IRAM,
  author={Patterson, D.A.},
  booktitle={1999 International Symposium on VLSI Technology, Systems, and Applications. Proceedings of Technical Papers. (Cat. No.99TH8453)}, 
  title={IRAM: a microprocessor for the post-PC era}, 
  year={1999},
  volume={},
  number={},
  pages={39-41},
  doi={10.1109/VTSA.1999.785994}}

@INPROCEEDINGS{1992-Dally-MDP,
  author={Dally, W.J. and Chien, A. and Fiske, J.A.S. and Fyler, G. and Horwat, W. and Keen, J.S. and Lethin, R.A. and Noakes, M. and Nuth, P.R. and Wills, D.S.},
  booktitle={Proceedings 1992 IEEE International Conference on Computer Design: VLSI in Computers \& Processors}, 
  title={The Message Driven Processor: an integrated multicomputer processing element}, 
  year={1992},
  volume={},
  number={},
  pages={416-419},
  doi={10.1109/ICCD.1992.276304}}

@article{1993-Noakes-MDP,
author = {Noakes, Michael D. and Wallach, Deborah A. and Dally, William J.},
title = {The J-machine multicomputer: an architectural evaluation},
year = {1993},
issue_date = {May 1993},
publisher = {Association for Computing Machinery},
address = {New York, NY, USA},
volume = {21},
number = {2},
issn = {0163-5964},
url = {https://doi.org/10.1145/173682.165158},
doi = {10.1145/173682.165158},
journal = {SIGARCH Comput. Archit. News},
month = {may},
pages = {224–235},
numpages = {12}
}

@inproceedings{1987-Dally-Jmachine,
author = {Dally, W. J. and Chao, L. and Chien, A. and Hassoun, S. and Horwat, W. and Kaplan, J. and Song, P. and Totty, B. and Wills, S.},
title = {Architecture of a message-driven processor},
year = {1987},
isbn = {0818607769},
publisher = {Association for Computing Machinery},
address = {New York, NY, USA},
url = {https://doi.org/10.1145/30350.30372},
doi = {10.1145/30350.30372},
booktitle = {Proceedings of the 14th Annual International Symposium on Computer Architecture},
pages = {189–196},
numpages = {8},
location = {Pittsburgh, Pennsylvania, USA},
series = {ISCA '87}
}

@INPROCEEDINGS{1990-Blank-MASPAR,
  author={Blank, T.},
  booktitle={Digest of Papers Compcon Spring '90. Thirty-Fifth IEEE Computer Society International Conference on Intellectual Leverage}, 
  title={The MasPar MP-1 architecture}, 
  year={1990},
  volume={},
  number={},
  pages={20-24},
  doi={10.1109/CMPCON.1990.63648}}

@techreport{1990-Maspar-Fortran,
  author      = "MasPar",
  title       = "MasPar Application Language (MPL) Reference Manual,",
  institution = "MasPar",
  year        = "1990",
  type        = "",
  number      = "Pub-9302-0000",
  address     = "",
  month       = "Jan",
  note        = ""
}

@INPROCEEDINGS{1998-Nunomura-M32,
  author={Nunomura, Y. and Shimizu, T. and Saitoh, K. and Tsuchihashi, K.},
  booktitle={Proceedings of the 1998 IEEE International Conference on Acoustics, Speech and Signal Processing, ICASSP '98 (Cat. No.98CH36181)}, 
  title={Multimedia applications of microprocessor with embedded DRAM}, 
  year={1998},
  volume={5},
  number={},
  pages={3157-3160 vol.5},
  doi={10.1109/ICASSP.1998.678196}}

@INPROCEEDINGS{1996-Shimizu-M32,
  author={Shimizu, T. and Korematu, J. and Satou, M. and Kondo, H. and Iwata, S. and Sawai, K. and Okumura, N. and Ishimi, K. and Nakamoto, Y. and Kumanoya, M. and Dosaka, K. and Yamazaki, A. and Ajioka, Y. and Tsubota, H. and Nunomura, Y. and Urabe, T. and Hinata, J. and Saitoh, K.},
  booktitle={1996 IEEE International Solid-State Circuits Conference. Digest of TEchnical Papers, ISSCC}, 
  title={A multimedia 32 b RISC microprocessor with 16 Mb DRAM}, 
  year={1996},
  volume={},
  number={},
  pages={216-217},
  doi={10.1109/ISSCC.1996.488577}}

@ARTICLE{1997-Nunomura-M32,
  author={Nunomura, Y. and Shimizu, T. and Tomisawa, O.},
  journal={IEEE Micro}, 
  title={M32R/D-integrating DRAM and microprocessor}, 
  year={1997},
  volume={17},
  number={6},
  pages={40-48},
  doi={10.1109/40.641595}}

@ARTICLE{1998-Shimizu-M32RxD,
  author={Toru Shimizu},
  journal={Hot Chips 10}, 
  title={M32Rx/D - A Single Chip
Microcontroller with A High Capacity
4MB Internal DRAM}, 
  year={1998},
  volume={},
  number={},
  pages={},
}

@INPROCEEDINGS{2020-WAN-NN-RRAM,
  author={Wan, Weier and Kubendran, Rajkumar and Eryilmaz, S. Burc and Zhang, Wenqiang and Liao, Yan and Wu, Dabin and Deiss, Stephen and Gao, Bin and Raina, Priyanka and Joshi, Siddharth and Wu, Huaqiang and Cauwenberghs, Gert and Wong, H.-S. Philip},
  booktitle={2020 IEEE International Solid-State Circuits Conference - (ISSCC)}, 
  title={33.1 A 74 TMACS/W CMOS-RRAM Neurosynaptic Core with Dynamically Reconfigurable Dataflow and In-situ Transposable Weights for Probabilistic Graphical Models}, 
  year={2020},
  volume={},
  number={},
  pages={498-500},
  doi={10.1109/ISSCC19947.2020.9062979}}

@INPROCEEDINGS{2023-Song-Refloat,
  author={Song, Linghao and Chen, Fan and Li, Hai and Chen, Yiran},
  booktitle={SC23: International Conference for High Performance Computing, Networking, Storage and Analysis}, 
  title={Refloat: Low-Cost Floating-Point Processing in ReRAM for Accelerating Iterative Linear Solvers}, 
  year={2023},
  volume={},
  number={},
  pages={1-15},
  doi={}}

@INPROCEEDINGS{2023-Fang-NN-Rel-PSP,
  author={Fang, Wenbing and Xuan, Zihao and Chen, Song and Kang, Yi},
  booktitle={2023 IEEE Biomedical Circuits and Systems Conference (BioCAS)}, 
  title={An 1.38nJ/Inference Clock-Free Mixed-Signal Neuromorphic Architecture Using ReL-PSP Function and Computing-in-Memory}, 
  year={2023},
  volume={},
  number={},
  pages={1-5},
  doi={10.1109/BioCAS58349.2023.10388821}}

@INPROCEEDINGS {2024-Andrulis-CIMLoop,
author = { Andrulis, Tanner and Emer, Joel S. and Sze, Vivienne },
booktitle = { 2024 IEEE International Symposium on Performance Analysis of Systems and Software (ISPASS) },
title = {{ CiMLoop: A Flexible, Accurate, and Fast Compute-In-Memory Modeling Tool }},
year = {2024},
volume = {},
ISSN = {},
pages = {10-23},
doi = {10.1109/ISPASS61541.2024.00012},
url = {https://doi.ieeecomputersociety.org/10.1109/ISPASS61541.2024.00012},
publisher = {IEEE Computer Society},
address = {Los Alamitos, CA, USA},
month =May}

@ARTICLE{2024-Chen-NN-RRAM,
  author={Chen, Yiyang and Han, Lixia and Zhang, Yizhou and Zhou, Zheng and Huang, Peng and Li, Can and Liu, Lifeng and Kang, Jinfeng},
  journal={IEEE Electron Device Letters}, 
  title={Inner Product Accelerating Scheme Based on RRAM Array for Attention-Mechanism Neural Network}, 
  year={2024},
  volume={45},
  number={3},
  pages={360-363},
  doi={10.1109/LED.2024.3352087}}

@INPROCEEDINGS{2022-Kim-NANDPIM,
  author={Kim, HyunWoo and Baek, Seungwon and Song, Jaehong and Song, Taigon},
  booktitle={2022 19th International SoC Design Conference (ISOCC)}, 
  title={A Novel Processing Unit and Architecture for Process-In Memory (PIM) in NAND Flash Memory}, 
  year={2022},
  volume={},
  number={},
  pages={127-128},
  doi={10.1109/ISOCC56007.2022.10031375}}

@INPROCEEDINGS {2024-Chen-PointCIM,
author = { Chen, Xuan-Jun and Chen, Han-Ping and Yang, Chia-Lin },
booktitle = { 2024 57th IEEE/ACM International Symposium on Microarchitecture (MICRO) },
title = {{ PointCIM: A Computing-in-Memory Architecture for Accelerating Deep Point Cloud Analytics }},
year = {2024},
volume = {},
ISSN = {},
pages = {1309-1322},
doi = {10.1109/MICRO61859.2024.00097},
url = {https://doi.ieeecomputersociety.org/10.1109/MICRO61859.2024.00097},
publisher = {IEEE Computer Society},
address = {Los Alamitos, CA, USA},
month =Nov}

@ARTICLE{2024-Shin-Piccolo,
  author={Shin, Changmin and Kwon, Taehee and Song, Jaeyong and Ju, Jae Hyung and Liu, Frank and Choi, Yeonkyu and Lee, Jinho},
  journal={IEEE Computer Architecture Letters}, 
  title={A Case for In-Memory Random Scatter-Gather for Fast Graph Processing}, 
  year={2024},
  volume={23},
  number={1},
  pages={73-77},
  doi={10.1109/LCA.2024.3376680}}

@inproceedings{2005-Thoziyoor-PIMLite,
author = {Thoziyoor, Shyamkumar and Brockman, Jay and Rinzler, Daniel},
title = {PIM lite: a multithreaded processor-in-memory prototype},
year = {2005},
isbn = {1595930574},
publisher = {Association for Computing Machinery},
address = {New York, NY, USA},
url = {https://doi.org/10.1145/1057661.1057678},
doi = {10.1145/1057661.1057678},
booktitle = {Proceedings of the 15th ACM Great Lakes Symposium on VLSI},
pages = {64–69},
numpages = {6},
location = {Chicago, Illinois, USA},
series = {GLSVLSI '05}
}

@inproceedings{2004-Brockman-PIMLite,
author = {Brockman, Jay B. and Thoziyoor, Shyamkumar and Kuntz, Shannon K. and Kogge, Peter M.},
title = {A low cost, multithreaded processing-in-memory system},
year = {2004},
isbn = {159593040X},
publisher = {Association for Computing Machinery},
address = {New York, NY, USA},
url = {https://doi.org/10.1145/1054943.1054946},
doi = {10.1145/1054943.1054946},
booktitle = {Proceedings of the 3rd Workshop on Memory Performance Issues: In Conjunction with the 31st International Symposium on Computer Architecture},
pages = {16–22},
numpages = {7},
location = {Munich, Germany},
series = {WMPI '04}
}

@TECHREPORT{2003-Brockman-PIMLite,
  author = {Jay Brockman and Peter M. Kogge and Shyamkumar Thoziyoor and Edward
	Kang},
  title = {{{PIM} Lite: On} the road towards relentless multi-threading in massively
	parallel systems},
  institution = "Univ. of Notre Dame, CSE Dept.",
  year = {2003},
 url = {\url{https://citeseerx.ist.psu.edu/document?repid=rep1&type=pdf&doi=c4bbfa7e7519337824709da275cf3f6c26314475}}}

@techreport{2003-Brockman-PIMLite-Asm,
  author      = "Jay B. Brockman",
  title       = "PIM Lite Architecture and Assembly Language Programming Guide",
  institution = "Univ. of Notre Dame, CSE Dept.",
  year        = "2003",
  type        = "",
  number      = "",
  address     = "",
  month       = "July",
  note        = ""
}

@techreport{2003-Brockman-PIMLite-programming,
  author      = "Jay B. Brockman",
  title       = "Programming PIM Lite",
  institution = "Univ. of Notre Dame, CSE Dept.",
  year        = "2003",
  type        = "",
  number      = "",
  address     = "",
  month       = "July",
  note        = ""
}

@ARTICLE{1995-Kogge-RTAIS,
  author = {P Kogge and T. Giambra and H Sasnowitz},
  title = {{RTAIS}: An embedded parallel processor for real-time decision aiding},
  journal = {NAECON},
  year = {1995},
howpublished = {\url{https://ieeexplore.ieee.org/document/521976}}
}

@ARTICLE{2011-Yu-ReRAM,
  author={Yu, Shimeng and Wu, Yi and Jeyasingh, Rakesh and Kuzum, Duygu and Wong, H.-S. Philip},
  journal={IEEE Transactions on Electron Devices}, 
  title={An Electronic Synapse Device Based on Metal Oxide Resistive Switching Memory for Neuromorphic Computation}, 
  year={2011},
  volume={58},
  number={8},
  pages={2729-2737},
  doi={10.1109/TED.2011.2147791}}

@ARTICLE{2022-Kubendran-ReRAM,
  author={Wan, W and Kubendran, R and Schaefer, C and Eryilmaz, SB and Zhang, W and Wu, D and Deiss, S and Raina, P and Qian, H and Gao, B and Joshi, S and Wu, H and Wong, HP and Cauwenberghs, G},
  journal={Nature}, 
  title={A compute-in-memory chip based on resistive random-access memory}, 
  year={2022},
  volume={608},
  number={},
  pages={504-512},
  doi={10.1038/s41586-022-04992-8}}

@INPROCEEDINGS{2000-Mai-SmartMemories,
  author={Mai, K. and Paaske, T. and Jayasena, N. and Ho, R. and Dally, W.J. and Horowitz, M.},
  booktitle={Proceedings of 27th International Symposium on Computer Architecture (IEEE Cat. No.RS00201)}, 
  title={Smart Memories: a modular reconfigurable architecture}, 
  year={2000},
  volume={},
  number={},
  pages={161-171},
  doi={10.1109/ISCA.2000.854387}}

@INPROCEEDINGS{1993-Kessler-T3D,
  author={Kessler, R.E. and Schwarzmeier, J.L.},
  booktitle={Digest of Papers. Compcon Spring}, 
  title={Cray T3D: a new dimension for Cray Research}, 
  year={1993},
  volume={},
  number={},
  pages={176-182},
  doi={10.1109/CMPCON.1993.289660}}

@article{1995-Gokhale-Terasys,
author = {Gokhale, Maya and Holmes, Bill and Iobst, Ken},
year = {1995},
month = {05},
pages = {23 - 31},
title = {Processing in Memory: The Terasys Massively Parallel PIM Array},
volume = {28},
journal = {Computer},
doi = {10.1109/2.375174}
}

@INPROCEEDINGS{2000-Margolus-SpaceRAM,
  author={Margolus, N.},
  booktitle={Proceedings of 27th International Symposium on Computer Architecture (IEEE Cat. No.RS00201)}, 
  title={An embedded DRAM architecture for large-scale spatial-lattice computations}, 
  year={2000},
  volume={},
  number={},
  pages={149-160},
  doi={10.1145/339647.339672}}

@inproceedings{1974-Batcher-Staran,
author = {Batcher, Kenneth E.},
title = {STARAN parallel processor system hardware},
year = {1974},
isbn = {9781450379205},
publisher = {Association for Computing Machinery},
address = {New York, NY, USA},
url = {https://doi.org/10.1145/1500175.1500260},
doi = {10.1145/1500175.1500260},
booktitle = {Proceedings of the May 6-10, 1974, National Computer Conference and Exposition},
pages = {405–410},
numpages = {6},
location = {Chicago, Illinois},
series = {AFIPS '74}
}

@ARTICLE{1977-Batcher-Staran-memory,
  author={Batcher, Kenneth E.},
  journal={IEEE Transactions on Computers}, 
  title={The Multidimensional Access Memory in STARAN}, 
  year={1977},
  volume={C-26},
  number={2},
  pages={174-177},
  doi={10.1109/TC.1977.5009297}}

@article{1976-Batcher-Staran-network,
author = {Batcher, Kenneth},
year = {1976},
month = {01},
pages = {},
title = {The flip network in staran},
journal = {IEEE Transactions on Computers - TC}
}

@ARTICLE{1982-Batcher-ASPRO-MPP,
  author={Batcher},
  journal={IEEE Transactions on Computers}, 
  title={Bit-Serial Parallel Processing Systems}, 
  year={1982},
  volume={C-31},
  number={5},
  pages={377-384},
  doi={10.1109/TC.1982.1676015}}

@INPROCEEDINGS{2003-Kirsch-Yukon,
  author={Kirsch, G.},
  booktitle={Proceedings International Parallel and Distributed Processing Symposium}, 
  title={Active memory: Micron's Yukon}, 
  year={2003},
  volume={},
  number={},
  pages={11 pp.-},
  doi={10.1109/IPDPS.2003.1213195}}

@ARTICLE{2022-Coluccio-Hybrid-SIMD,
  author={Coluccio, Andrea and Casale, Umberto and Guastamacchia, Angela and Turvani, Giovanna and Vacca, Marco and Roch, Massimo Ruo and Zamboni, Maurizio and Graziano, Mariagrazia},
  journal={IEEE Transactions on Computers}, 
  title={Hybrid-SIMD: A Modular and Reconfigurable Approach to Beyond von Neumann Computing}, 
  year={2022},
  volume={71},
  number={9},
  pages={2287-2299},
  doi={10.1109/TC.2021.3127354}}

@INPROCEEDINGS{2003-Rodrigues-PIM-MPI,
  author={Arun Rodrigues and Richard Murphy and Peter Kogge and Jay Brockman and Ron Brightwell and Keith Underwood},
  booktitle={2003 Proceedings IEEE International Conference on Cluster Computing}, 
  title={Implications of a PIM architectural model for MPI}, 
  year={2003},
  volume={},
  number={},
  pages={259-268},
  doi={10.1109/CLUSTR.2003.1253323}}

@ARTICLE{2007-McDonald-Transactional,
  author={McDonald, Austen and Carlstrom, Brian D. and Chung, JaeWoong and Minh, Chi Cao and Chafi, Hassan and Kozyrakis, Christos and Olukotun, Kunle},
  journal={IEEE Micro}, 
  title={Transactional Memory: The Hardware-Software Interface}, 
  year={2007},
  volume={27},
  number={1},
  pages={67-76},
  doi={10.1109/MM.2007.26}}

@ARTICLE{1999-Birnbaum-VSIA,
  author={Birnbaum, M. and Sachs, H.},
  journal={Computer}, 
  title={How VSIA answers the SOC dilemma}, 
  year={1999},
  volume={32},
  number={6},
  pages={42-50},
  doi={10.1109/2.769442}}

@INPROCEEDINGS{1996-Nowatzyk-wall,
  author={Nowatzyk, A. and Fong Pong and Saulsbury, A.},
  booktitle={23rd Annual International Symposium on Computer Architecture (ISCA'96)}, 
  title={Missing the Memory Wall: The Case for Processor/Memory Integration}, 
  year={1996},
  volume={},
  number={},
  pages={90-90},
  doi={10.1145/232973.232984}}

@INPROCEEDINGS{1998-Oskin-ActivePages,
  author={Oskin, M. and Chong, F.T. and Sherwood, T.},
  booktitle={Proceedings. 25th Annual International Symposium on Computer Architecture (Cat. No.98CB36235)}, 
  title={Active Pages: a computation model for intelligent memory}, 
  year={1998},
  volume={},
  number={},
  pages={192-203},
  doi={10.1109/ISCA.1998.694774}}

@INPROCEEDINGS{2022-Chen-RNASeq,
  author={Chen, Liang-Chi and Yu, Shu-Qi and Ho, Chien-Chung and Chang, Yuan-Hao and Chang, Da-Wei and Wang, Wei-Chen and Chang, Yu-Ming},
  booktitle={2022 IEEE 11th Non-Volatile Memory Systems and Applications Symposium (NVMSA)}, 
  title={RNA-seq Quantification on Processing in memory Architecture: Observation and Characterization}, 
  year={2022},
  volume={},
  number={},
  pages={26-32},
  doi={10.1109/NVMSA56066.2022.00014}}

@INPROCEEDINGS{1997-Burger-WhyPIM,
  author={Burger, D. and Goodman, J.R.},
  booktitle={Proceedings Innovative Architecture for Future Generation High-Performance Processors and Systems}, 
  title={Memory-centric architectures: why and perhaps what}, 
  year={1997},
  volume={},
  number={},
  pages={92-},
  doi={10.1109/IWIA.1997.670413}}

@INPROCEEDINGS{1998-Rixner-MediaPIM,
  author={Rixner, S. and Dally, W.J. and Kapasi, U.J. and Khailany, B. and Lopez-Lagunas, A. and Mattson, P.R. and Owens, J.D.},
  booktitle={Proceedings. 31st Annual ACM/IEEE International Symposium on Microarchitecture}, 
  title={A bandwidth-efficient architecture for media processing}, 
  year={1998},
  volume={},
  number={},
  pages={3-13},
  doi={10.1109/MICRO.1998.742118}}

@ARTICLE{2019-Zabihi-Spintronics,
  author={Zabihi, Masoud and Chowdhury, Zamshed Iqbal and Zhao, Zhengyang and Karpuzcu, Ulya R. and Wang, Jian-Ping and Sapatnekar, Sachin S.},
  journal={IEEE Transactions on Computers}, 
  title={In-Memory Processing on the Spintronic CRAM: From Hardware Design to Application Mapping}, 
  year={2019},
  volume={68},
  number={8},
  pages={1159-1173},
  doi={10.1109/TC.2018.2858251}}

@ARTICLE{2017-Sato-RSFQ,
  author={Sato, Ryo and Hatanaka, Yuki and Ando, Yuki and Tanaka, Masamitsu and Fujimaki, Akira and Takagi, Kazuyoshi and Takagi, Naofumi},
  journal={IEEE Transactions on Applied Superconductivity}, 
  title={High-Speed Operation of Random-Access-Memory-Embedded Microprocessor With Minimal Instruction Set Architecture Based on Rapid Single-Flux-Quantum Logic}, 
  year={2017},
  volume={27},
  number={4},
  pages={1-5},
  doi={10.1109/TASC.2016.2642049}}

@ARTICLE{1970-Stone-LogicInMemory,
  author={Stone, Harold S.},
  journal={IEEE Transactions on Computers}, 
  title={A Logic-in-Memory Computer}, 
  year={1970},
  volume={C-19},
  number={1},
  pages={73-78},
  doi={10.1109/TC.1970.5008902}}

@INPROCEEDINGS{1992-Elliot-CompRAM,
  author={Elliott, D.G. and Snelgrove, W.M. and Stumm, M.},
  booktitle={1992 Proceedings of the IEEE Custom Integrated Circuits Conference}, 
  title={Computational Ram: A Memory-simd Hybrid And Its Application To Dsp}, 
  year={1992},
  volume={},
  number={},
  pages={30.6.1-30.6.4},
  doi={10.1109/CICC.1992.591879}}

@ARTICLE{1969-Kautz-cellular,
  author={Kautz, W.H.},
  journal={IEEE Transactions on Computers}, 
  title={Cellular Logic-in-Memory Arrays}, 
  year={1969},
  volume={C-18},
  number={8},
  pages={719-727},
  doi={10.1109/T-C.1969.222754}}

@ARTICLE{2025-kogge-migrating,
  author={Peter M. Kogge and et al.},
  journal={To be published}, 
  title={The Case for Migrating Threads}, 
  year={2025},
  volume={},
  number={},
  pages={},
  doi={}}

@inproceedings{2018-Mohammedali-acc-in-cloud,
author = {Mohammedali, Noor and Agyeman, Michael Opoku},
title = {A Study of Reconfigurable Accelerators for Cloud Computing},
year = {2018},
isbn = {9781450366281},
publisher = {Association for Computing Machinery},
address = {New York, NY, USA},
url = {https://doi.org/10.1145/3284557.3284563},
doi = {10.1145/3284557.3284563},
booktitle = {Proceedings of the 2nd International Symposium on Computer Science and Intelligent Control},
articleno = {17},
numpages = {5},
location = {Stockholm, Sweden},
series = {ISCSIC '18}
}

@article{2022-Asufuzzaman-Survey,
title = {A survey on processing-in-memory techniques: Advances and challenges},
journal = {Memories - Materials, Devices, Circuits and Systems},
volume = {4},
pages = {100022},
year = {2023},
issn = {2773-0646},
doi = {https://doi.org/10.1016/j.memori.2022.100022},
url = {https://www.sciencedirect.com/science/article/pii/S2773064622000160},
author = {Kazi Asifuzzaman and Narasinga Rao Miniskar and Aaron R. Young and Frank Liu and Jeffrey S. Vetter}
}

@inproceedings{2019-Imani-DigitalPIM,
author = {Imani, Mohsen and Gupta, Saransh and Rosing, Tajana},
title = {Digital-based processing in-memory: a highly-parallel accelerator for data intensive applications},
year = {2019},
isbn = {9781450372060},
publisher = {Association for Computing Machinery},
address = {New York, NY, USA},
url = {https://doi.org/10.1145/3357526.3357551},
doi = {10.1145/3357526.3357551},
booktitle = {Proceedings of the International Symposium on Memory Systems},
pages = {38–40},
numpages = {3},
location = {Washington, District of Columbia, USA},
series = {MEMSYS '19}
}

@ARTICLE{2022-Leitersdorf-MultPIM,
  author={Leitersdorf, Orian and Ronen, Ronny and Kvatinsky, Shahar},
  journal={IEEE Transactions on Circuits and Systems II: Express Briefs}, 
  title={MultPIM: Fast Stateful Multiplication for Processing-in-Memory}, 
  year={2022},
  volume={69},
  number={3},
  pages={1647-1651},
  doi={10.1109/TCSII.2021.3118215}}

@INPROCEEDINGS{2019-Peng-RRAM-RNN,
  author={Peng, Xiaochen and Liu, Rui and Yu, Shimeng},
  booktitle={2019 IEEE International Symposium on Circuits and Systems (ISCAS)}, 
  title={Optimizing Weight Mapping and Data Flow for Convolutional Neural Networks on RRAM Based Processing-In-Memory Architecture}, 
  year={2019},
  volume={},
  number={},
  pages={1-5},
  doi={10.1109/ISCAS.2019.8702715}}

@ARTICLE{2018-Long-RRAM-RNN,
  author={Long, Yun and Na, Taesik and Mukhopadhyay, Saibal},
  journal={IEEE Transactions on Very Large Scale Integration (VLSI) Systems}, 
  title={ReRAM-Based Processing-in-Memory Architecture for Recurrent Neural Network Acceleration}, 
  year={2018},
  volume={26},
  number={12},
  pages={2781-2794},
  doi={10.1109/TVLSI.2018.2819190}}

@INPROCEEDINGS{2021-Anni-ML,
  author={Lu, Anni and Peng, Xiaochen and Luo, Yandong and Huang, Shanshi and Yu, Shimeng},
  booktitle={2021 Design, Automation and Test in Europe Conference and Exhibition (DATE)}, 
  title={A Runtime Reconfigurable Design of Compute-in-Memory based Hardware Accelerator}, 
  year={2021},
  volume={},
  number={},
  pages={932-937},
  doi={10.23919/DATE51398.2021.9474156}}

@INPROCEEDINGS{2021-Kazemi-inferencing,
  author={Kazemi, Arman and Sharifi, Mohammad Mehdi and Zou, Zhuowen and Niemier, Michael and Hu, X. Sharon and Imani, Mohsen},
  booktitle={2021 IEEE/ACM International Symposium on Low Power Electronics and Design (ISLPED)}, 
  title={MIMHD: Accurate and Efficient Hyperdimensional Inference Using Multi-Bit In-Memory Computing}, 
  year={2021},
  volume={},
  number={},
  pages={1-6},
  doi={10.1109/ISLPED52811.2021.9502498}}

@ARTICLE{2021-Roy-PIM-DRAM,
  author={Roy, Sourjya and Ali, Mustafa and Raghunathan, Anand},
  journal={IEEE Journal on Emerging and Selected Topics in Circuits and Systems}, 
  title={PIM-DRAM: Accelerating Machine Learning Workloads Using Processing in Commodity DRAM}, 
  year={2021},
  volume={11},
  number={4},
  pages={701-710},
  doi={10.1109/JETCAS.2021.3127517}}

@ARTICLE{2019-Long-ReRAM-DNN,
  author={Long, Yun and Kim, Daehyun and Lee, Edward and Saha, Priyabrata and Mudassar, Burhan Ahmad and She, Xueyuan and Khan, Asif Islam and Mukhopadhyay, Saibal},
  journal={IEEE Journal on Exploratory Solid-State Computational Devices and Circuits}, 
  title={A Ferroelectric FET-Based Processing-in-Memory Architecture for DNN Acceleration}, 
  year={2019},
  volume={5},
  number={2},
  pages={113-122},
  doi={10.1109/JXCDC.2019.2923745}}

@ARTICLE{2019-Lee-DRAM-PIM,
  author={Lee, Won Jun and Kim, Chang Hyun and Paik, Yoonah and Park, Jongsun and Park, Il and Kim, Seon Wook},
  journal={IEEE Access}, 
  title={Design of Processing-“Inside”-Memory Optimized for DRAM Behaviors}, 
  year={2019},
  volume={7},
  number={},
  pages={82633-82648},
  doi={10.1109/ACCESS.2019.2924240}}

@INPROCEEDINGS{2019-Zhao-BNN-PIMs,
  author={Zhao, Yinglin and Yang, Jianlei and Jia, Xiaotao and Wang, Xueyan and Wang, Zhaohao and Kang, Wang and Zhang, Youguang and Zhao, Weisheng},
  booktitle={2019 IEEE Computer Society Annual Symposium on VLSI (ISVLSI)}, 
  title={Exploiting Near-Memory Processing Architectures for Bayesian Neural Networks Acceleration}, 
  year={2019},
  volume={},
  number={},
  pages={203-206},
  doi={10.1109/ISVLSI.2019.00045}}

@misc{2023-upmem-abumpimp,
      title={ABUMPIMP 2023}, 
      author={Upmem},
      year={2023},
      url={https://www.overleaf.com/project/67a4cb2713ff3659f2e5a61b}, 
}

\printindex

\end{document}